\documentclass[12pt,a4paper]{article}
\usepackage{epsfig}
\begin{document}
\title{Search for the vector-like top quark $T$ at the future
linac-ring type $ep$ collider}

\author{ Chong-Xing Yue, Feng Zhang,  Wei Wang\\
{\small  Department of Physics, Liaoning Normal University, Dalian
116029, China}\thanks{E-mail:cxyue@lnnu.edu.cn}\\}
\date{\today}

\maketitle
\begin{abstract}

The little Higgs  models typically contain a new vector-like top
quark $T$, which plays a key role in breaking the electroweak
symmetry. In the context of the littlest Higgs (LH) model, we
study single production of this kind of new particle via the
process $ep\rightarrow eb\rightarrow \nu_{e}T$ in the future
linac-ring type $ep$ collider (LC$\bigotimes LHC$). We find that
the production cross section is in the range of $1.2\times
10^{-4}$--- 0.48 $pb$ at the $LC\bigotimes LHC$ with
$\sqrt{s}=3.7$ $TeV$.
 \vspace{1cm}

\hspace{-0.5cm}PACS number(s):12.60.Cn, 13.60.Fz, 14.65.Ha

\end{abstract}

\newpage
Little Higgs models[1,2,3] were recently proposed as a kind of
models of electroweak symmetry breaking(EWSB), which can solve the
hierarchy problem by protecting the Higgs mass  from quadratically
divergent at one-loop order and thus can be regarded as one of the
important candidates of the new physics beyond the standard
model(SM). The key feature of this kind of models is that the
Higgs boson is a pseudo-Goldstone boson of a global symmetry
breaking at a scale $\Lambda\sim10TeV$, so that the Higgs boson
mass can be as light as $O(100GeV)$. The light Higgs boson mass is
protected from the one-loop quadratic divergence by introducing a
few new particles with the same statistics as the corresponding
$SM$ particles. The new heavy gauge bosons cancel the one-loop
quadratic divergence generated by the $SM$ gauge boson $W$ and $Z$
loops, new heavy scalars cancel that by the Higgs
self-interaction, while the new vector-like top quark $T$ cancels
that by the top quark Yukawa interactions. Furthermore, these new
particles might produce characteristic signatures at the present
and future collider experiments[4,5,6]. Certainly, these new
particles can generate significant corrections to some observables
and thus the precision measurement data can give severe
constraints on this kind of models[4,7,8].

In little Higgs models, $EWSB$ generally results from the coupling
of the Higgs boson to an independent sector containing the $SM$
top quark $t$ and the new vector-like top quark $T$. This kind of
models provide a natural mechanism of $EWSB$ associated with the
large value of the top quark Yukawa coupling, in which the new
vector-like top quark $T$ plays a key role in breaking the
electroweak symmetry. Thus, studying the possible signatures of
the new particle $T$ at present and future high energy colliders
would provide crucial information for $EWSB$ and future test the
little Higgs models.

To avoid the fine tuning problem and produce a suitable Higgs
mass, the mass of the new vector-like top quark $T$ should not be
too large. Considering this reason and the precision electroweak
constraints on little Higgs models, if the little Higgs models are
correct, the $T$ mass $M_{T}$ should be about $2TeV$[1]. In this
case, the new particle $T$ can be produced at the $LHC$ via two
mechanism: $QCD$ pair production via the processes $gg\rightarrow
T\overline{T}$ and $q\overline{q}\rightarrow T\overline{T}$;
single production via $W$ exchange process $qb\rightarrow q'T$.
Due to the large $T$ mass $M_{T}$, the later process dominates
over the $QCD$ pair production process. It has been shown that the
new heavy top quark $T$ mass $M_{T}$ can be explored up to about
$2.5TeV$ via the $W$ exchange process $qb\rightarrow q'T$[4,5].

Although the linac-ring type $ep$ collider($LC\otimes LHC$) with
the centre-of-mass(c.m.) energy $\sqrt{s}=3.7TeV$ and the integral
luminosity $\pounds_{int}\approx 100pb^{-1}$ has a lower
luminosity, it can provide better conditions for studying a lot of
phenomena comparing to $LC$ due to the higher c. m. energy and to
$LHC$ due to more clear environment[9]. Thus, it can be used to
detect the possible signals of the new heavy particles. For
example, Ref.[10] has recently discussed the production of excited
neutrinos at this type of collider. In this letter, we will study
single production of the heavy vector-like top quark $T$ predicted
by the littlest Higgs($LH$) model[1] via the process $ep
\rightarrow eb\rightarrow \nu_{e}T$ and see whether it can be
detected in the future  $LC\otimes LHC$ with the c.m. energy
$\sqrt{s}=3.7TeV$ and the integral luminosity $\pounds
_{int}\approx 100pb^{-1}$.

As the simplest realization of the little Higgs idea, the $LH$
model[1] is a phenomenologically viable model and has almost all
of the essential features of the little Higgs models. So, in this
letter, we will give our results in the framework of the $LH$
model, although many alternatives have been proposed[2,3].
However, the presence of the heavy vector-like top quark $T$ is an
essential feature of this kind of models, thus our results might
be apply to other models.

In the $LH$ model, the couplings of the heavy vector-like top
quark $T$ to ordinary particles, which are related to our
calculation, can be written as[4]:
\begin{eqnarray}
g_{L}^{WTb}&=&\frac{e}{\sqrt{2}S_{W}}\frac{\nu}{f}x_{L},\hspace{1cm}g_{R}^{WTb}=0;\\
g_{L}^{W_{H}Tb}&=&-\frac{e}{\sqrt{2}S_{W}}\frac{\nu}{f}\frac{c}{s}x_{L},
\hspace{1cm}g_{R}^{W_{H}Tb}=0;\\
g_{L}^{W_{H}\nu e}&=&-\frac{e}{\sqrt{2}S_{W}}\frac{c}{s},\hspace{1cm}g_{R}^{W_{H}\nu e}=0;\\
g_{L}^{ZTt}&=&-\frac{ie}{2S_{W}C_{W}}\frac{\nu}{f}x_{L},\hspace{1cm}g_{R}^{ZTt}=0;\\
g_{L}^{HTt}&=&\frac{m_{t}}{\nu}\sqrt{\frac{x_{L}}{1-x_{L}}},
\hspace{1cm}g_{R}^{HTt}=\frac{m_{t}}{\nu}(1+x_{L})\frac{\nu}{f},
\end{eqnarray}
where $f$ is the scale parameter, $\nu\simeq 246 GeV$ is the
electroweak scale, $S_{W}=\sin \theta_{W}$, $\theta_{W}$ is the
Weinberg angle, and $c$ is the mixing parameter between
$SU(2)_{1}$ and $SU(2)_{2}$ gauge bosons with $ s^{2}= 1- c^{2}$.
$x_{L}$ is the mixing parameter between the $SM$ top quark $t$ and
the vector-like top quark $T$, which is defined as
$x_{L}=\lambda_{1}^{2}/(\lambda_{1}^{2}+\lambda_{2}^{2})$.
$\lambda_{1}$ and $\lambda_{2}$ are the Yukawa coupling
parameters. The mass of the heavy vector-like top quark $T$ can be
approximately written as:
\begin{equation}
M_{T}=\frac{m_{t}f}{\nu}\sqrt{\frac{1}{x_{L}(1-x_{L})}}
[1-\frac{\nu^{2}}{2f^{2}}x_{L}(1+x_{L})].
\end{equation}
In the following calculation, we will take $M_{T}$, $x_{L}$, and
$c$ as free  parameters.

The decay modes of the heavy vector-like top quark $T$ involving
the new particles $B_{H}$, $Z_{H}$, and $W_{H}$ might be
kinematically forbidden due to these new particles are heavy.
Thus, the dominate decay modes of $T$ are $tH$, $tZ$, and $bW$
with partial widths in the ratio $1:1:2$[4, 11]. At the order of
$\nu^{2}/f^{2}$, the total width of the new quark $T$ can be
written as:
\begin{equation}
\Gamma_{T}=\frac{M_{T}}{8\pi}(\frac{m_{t}}{\nu})^{2}\frac{x_{L}}{1-x_{L}}.
\end{equation}

In the $LH$ model, the number of up-type quarks is four and thus
the matrix relating the quark mass eigenstates with the weak
eigenstates becomes a $4\times 3$ matrix. Compared to the $CKM$
matrix in the SM, the extended  $CKM$ matrix has the fourth row
elements $V_{Td}$, $V_{Ts}$ and $V_{Tb}$. Thus, it is possible
that there are the decay channels $T\rightarrow ql\overline{\nu}$,
which $q$ is the down-type quark. However, their branching ratio
are very small. For example, Ref.$[12]$ has estimated the
branching ratio of the decay channel $T\rightarrow
bl\overline{\nu}$ and given $B_{r}$$(T\rightarrow bl\nu)$$\sim
1.2\times 10^{-3}\times (M_{T}/1TeV)^{4}$.

From above discussions, we can see that the heavy vector-like top
quark $T$ can be singly produced via the $t-$channel processes $ep
\rightarrow eb\rightarrow W^{*}\rightarrow \nu_{e}T$ and $ep
\rightarrow eb\rightarrow W_{H}^{*}\rightarrow \nu_{e}T$ in the
future $LC\otimes LHC$ with $\sqrt{s}=3.7TeV$. The relevant
Feynman diagrams are shown in Fig.1.

For the process $e(P_{e})+b(P_{b})\rightarrow
T(P_{T})+\nu_{e}(P_{\nu})$, we define the kinematical invariants
$t=(P_{T}-P_{b})^{2}$. The renormalization amplitude can be
written as:
\begin{eqnarray}
M^{T}&=&M_{W}^{T}+M_{W_{H}}^{T}\nonumber\\
&=&-\frac{e^{2}}{2S_{W}^{2}}\frac{\nu}{f}x_{L}\overline{u}_{\nu}(P_{\nu})
\frac{1-\gamma_{5}}{2}\gamma_{\mu}u_{e}(P_{e})
\frac{-i(g^{\mu\nu}-\frac{k^{\mu}k^{\nu}}{M_{W}^{2}})}{t-M_{W}^{2}+iM_{W}\Gamma_{W}}\nonumber\\
&&\hspace{0.5cm}\overline{u}_{T}(P_{T})\frac{1-\gamma_{5}}{2}\gamma_{\nu}u_{b}(P_{b})
-\frac{e^{2}}{2S_{W}^{2}}\frac{c^{2}}{s^{2}}\frac{\nu}{f}x_{L}\overline{u}_{\nu}
(P_{\nu})\frac{1-\gamma_{5}}{2}\gamma_{\mu}u_{e}(P_{e})\\
&&\hspace{0.5cm}\frac{-i(g^{\mu\nu}-\frac{k^{\mu}k^{\nu}}{M_{W_{H}}^{2}})}
{t-M_{W_{H}}^{2}+iM_{W_{H}}\Gamma_{W_{H}}}\overline{u}_{T}(P_{T})
\frac{1-\gamma_{5}}{2}\gamma_{\nu}u_{b}(P_{b}),\nonumber
\end{eqnarray}
where $\Gamma_{W}$ and $\Gamma_{W_{H}}$ are the total decay widths
of the $SM$ gauge boson $W$ and the new gauge boson $W_{H}$,
respectively. At the leading order, the decay modes of the new
gauge boson $W_{H}$ mainly contain $f\overline{f'}$ and $WH$. So,
the total decay width $\Gamma_{W_{H}}$ can be written as[11]:
\begin{equation}
\Gamma_{W_{H}}=\frac{\alpha_{e}}{96S_{W}^{2}}[\frac{96c^{2}}{s^{2}}
+\frac{C_{W}^{2}(c^{2}-s^{2})^{2}}{s^{2}c^{2}}]M_{W_{H}}.
\end{equation}
In above equation, we have ignored the contributions of the decay
modes $WZ$ and $WB_{H}$ to the total width $\Gamma_{W_{H}}$, which
are suppressed by a factor of $\nu^{4}/f^{4}$.

\begin{figure}[htb]
\vspace{-7.5cm}
\begin{center}
\epsfig{file=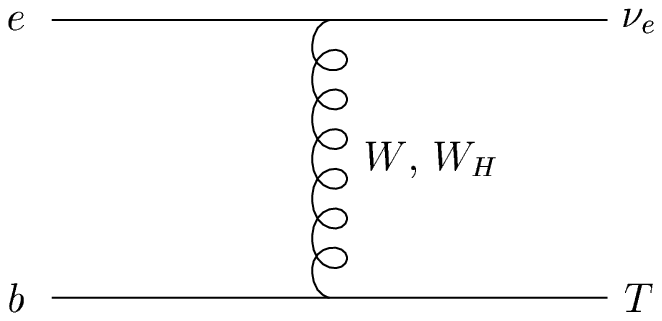,width=620pt,height=750pt} \vspace{-14cm}
\hspace{0.5cm} \caption{The Feynman diagrams of the $t-$channel
processes $eb\rightarrow W(W_{H})\rightarrow \nu_{e}T$.}
\label{ee}
\end{center}
\end{figure}

After calculating the cross section $\hat{\sigma}(\hat{s}$) of the
$t-$channel subprocess $eb\rightarrow\nu_{e} T$, the total cross
section $\sigma(s)$ of single $T$ production via the process
$ep\rightarrow eb\rightarrow \nu_{e}T$ at the future  $LC\otimes
LHC$ can be obtained by folding $\hat{\sigma}(\hat{s}$) with the
bottom-quark distribution function $f_{b}(x)$ in the proton:
\begin{equation}
\sigma(s)=\int_{x_{min}}^{1}f_{b}(x)\hat{\sigma}(x^{2}s)dx
\end{equation}
with $x_{min}=M_{T}/\sqrt{s}$. In our calculation, we will take
$CTEQ5$ parton distribution function[13] for $f_{b}(x)$.

From above equations, we can see that single $T$ production at the
$LC\otimes LHC$ comes from two processes: the $SM$ gauge boson $W$
exchange and the new gauge boson $W_{H}$ exchange. The
contributions of the former process $e^{-}b\rightarrow
W^{-}\rightarrow \nu_{e}T$ to single $T$ production mainly
dependent on the free parameters $M_{T}$ and $x_{L}$, while those
of the latter process $e^{-}b\rightarrow W_{H}^{-}\rightarrow
\nu_{e}T$ mainly dependent on the free parameters $M_{T}$,
$x_{L}$, $c$, and $M_{W_{H}}$. Taking into account the precision
electroweak constrains on the parameter space of the $LH$ model,
the free parameters $c$, $M_{W_{H}}$, $M_{T}$ and $x_{L}$ are
allowed in the ranges of $0\leq c\leq0.5$, $1TeV \leq
M_{W_{H}}\leq 3TeV$, and $0<x_{L}<1$ [8]. However, compared with
the contributions of $W$ exchange, the contributions of $W_{H}$
exchange to the cross section of single $T$ production is very
small. Thus, the cross section of single $T$ production is not
sensitive to the free parameters $c$ and $M_{W_{H}}$. So, in our
numerical estimation, we will take $c=0.3$ and $M_{W_{H}}$=$2TeV$.

\begin{figure}[htb]
\vspace{-0.2cm}
\begin{center}
\epsfig{file=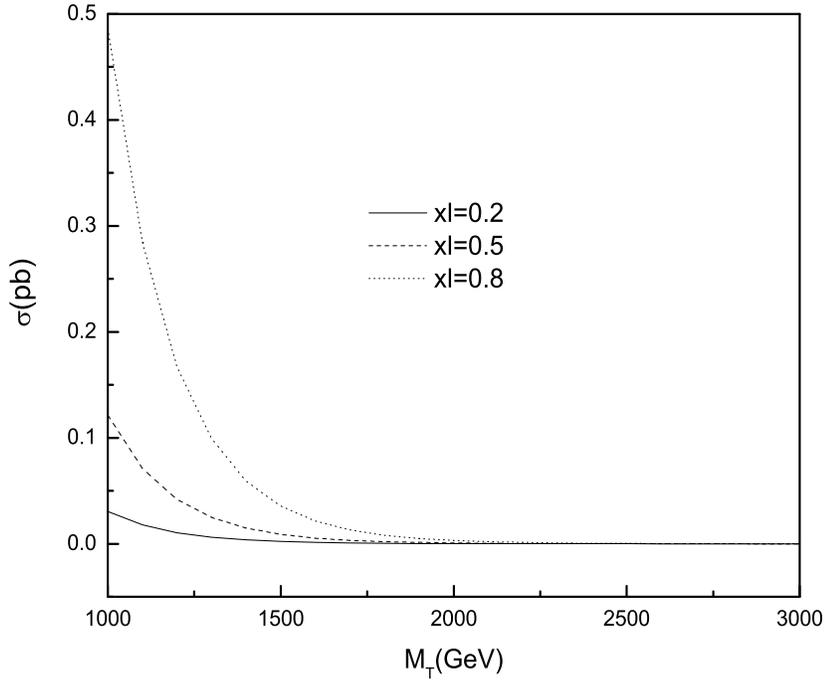,width=350pt,height=300pt} \vspace{-1cm}
\hspace{5mm} \caption{The cross section $\sigma(s)$ of single $T$
production as a function of $M_{T}$ for three values of the
parameter $x_{L}$.}
 \label{ee}
\end{center}
\end{figure}

In Fig.2, we plot the cross section of single $T$ production at
the $LC\otimes LHC$ with $\sqrt{s}=3.7$ $TeV$ as a function of the
$T$ quark mass $M_{T}$ for three values of the mixing parameter
$x_{L}$. One can see from Fig.2 that the cross section $\sigma
(s)$ increases as $M_{T}$ decreasing and $x_{L}$ increasing. For
$x_{L}$=0.5 and $1TeV\leq M_{T}\leq2.5TeV$, the value of the cross
section $\sigma(s)$ is in the range of $0.12pb\sim8
\times10^{-4}pb$. If we assume the integral luminosity of the
$LC\otimes LHC$ is $\pounds_{int}=100pb^{-1}$, then there will be
several and up to tens $\nu_{e}T$ events. Certainly, enhancing the
value of the integral luminosity of the $LC\otimes LHC$ can
largely increase the number of the $\nu_{e}T$ events.

At the leading order, the heavy vector-like top quark $T$
predicted by the $LH$ model mainly decays to the $tZ$, $tH$ and
$bW$ modes, which can provide characteristic signatures for the
discovery of the heavy vector-like quark $T$ in the future high
energy collider experiments. It has been shown that the signal of
the new vector-like quark $T$ might be detected via all of the
three decay modes in the future $LHC$ experiments[4,5]. Compared
with the $LHC$, the linac-ring type $ep$ collider, $LC\otimes
LHC$, has more clear environment. Furthermore, the cross sections
of single $T$ production at the $LHC$ and the $LC\otimes LHC$ are
at the same order of magnitude. Thus, the possible signal of the
vector-like top quark $T$ might be detected in the future
$LC\otimes LHC$ experiments. Certainly, more detailed analysis of
the $SM$ background would be needed to make a quantitative
conclusion for the $T$ observation.

\vspace{0.5cm} \noindent{\bf Acknowledgments}

F. Zhang would like to thank Bin Zhang for helpful discussions.
This work was supported in part by the National Natural Science
Foundation of China under the grant No.90203005 and No.10475037
and the Natural Science Foundation of the Liaoning Scientific
Committee(20032101).

\null
\end{document}